\documentclass[aps,prb,showpacs,twocolumn,amssymb]{revtex4}
\usepackage{graphicx}

\begin{document}

\draft

\title{Anomalous asymmetry of magnetoresistance in NbSe$_3$ single crystals}

\author{A.~A.~Sinchenko$^{a}$, Yu.~I. Latyshev$^{b}$, A.~P. Orlov$^{b}$, P.~Monceau$^{c}$}
\address{$^a$Moscow Engineering-Physics Institute, 115409
Moscow, Russia \\$^b$Institute of Radio Engineering and Electronics,
Russian Academy of Science, Moscow, 103907 Russia \\
$^c$Centre de Recherches sur les tres Basses Temperatures, BP 166,
38042 Grenoble, France}


\begin{abstract}

A pronounced asymmetry of magnetoresistance with respect to the
magnetic field direction is observed for NbSe$_3$ crystals placed in
a magnetic field perpendicular to their conducting planes. It is
shown that the effect persists in a wide temperature range and
manifests itself starting from a certain magnetic induction value
$B_0$, which at $T=4.2$ K corresponds to the transition to the
quantum limit, i.to the state where the Landay level splitting
exceeds the temperature.

\end{abstract}

\maketitle

The NbSe$_3$ material is one of the most popular
quasi-one-dimensional conductors with charge density waves (CDWs)
\cite{Grun}. The crystal lattice of NbSe$_3$ is monoclinic with
the b axis being parallel to the CDW chains and corresponding to
maximum conductivity. The anisotropy of conductivity in the
($b-c$) plane is $\sigma_b/\sigma_c \sim 10$, whereas the
conductivity ratio $\sigma_b/\sigma_a^*$ reaches a value of
$\sim10^4$ at low temperatures \cite{OngBrill,Latyshev2}. The
material experiences two Peierls transitions at the temperatures
$T_{p1}=$145 K and $T_{p2}=$59 K, below which the spectrum of
single-particle excitations develops energy gaps $\Delta_{p1}$ and
$\Delta_{p2}$ at the Fermi level. However, the electron spectrum
does not become completely dielectric. As a result of the
incomplete nesting, normal carriers, i.e., electrons and holes,
are retained in small pockets formed at the Fermi level
\cite{Grun}. The shape of the pockets was determined from the
angular dependences of Shubnikov–de Haas oscillations. According
to the data reported in Refs.
\onlinecite{Fleming78,Monceau78,Laborde87,Coleman90,sinch05}, the
Fermi surface areas that are not covered by the energy gap are
shaped as ellipsoids with a maximum axial ratio of $8-10$ and with
the major axes being parallel to the $c$ axis of the crystal. The
concentration of both types of carriers is $n\sim 10^{18}$
cm$^{-3}$ , their mobility at low temperature is $\mu\sim 10^6$
cm$^2$/V$\cdot$s (Refs.\onlinecite{OngBrill,Ong}), and the
effective mass is $m^*\sim10^{-1} m_e$
(Ref.\onlinecite{Coleman90}). Many of the experimental data can be
adequately explained under the assumption that, in NbSe$_3$ at low
temperatures, the two-dimensional nature of the electron spectrum
is realized \cite{sin03,Maki}. In view of the aforementioned
characteristics of the material, this suggests that the state of
the carriers should be close to that of a 2D electron gas. The
metal properties of NbSe$_3$ are retained down to the lowest
temperatures. Studies of this compound with an immobile CDW
revealed some unusual features of the transport properties due to
the carriers not condensed into the CDW. Primarily, these
properties include the effect of internal correlated interlayer
tunneling \cite{LatyshevPZ02,LatyshevJPA03} and the presence of
localized states within the Peierls energy gap
\cite{Latyshev05,Latyshev06}.

Magnetotransport properties of this material also exhibit unusual
behavior. When magnetic field is perpendicular to the $b$ axis of
the crystal, the magnetoresistance of NbSe$_3$ first rapidly
increases with magnetic field, then, at a certain field, its
growth becomes much slower. However, the resistance is not
saturated and, in high magnetic fields, the nonoscillating
component of magnetoresistance linearly varies with magnetic field
\cite{Coleman85,Richard87,Everson87,Tritt88,Monceau88}. In weak
magnetic fields, a quantum size effect is observed for magnetic
field orientations along the conducting layers (in the ($b–c$)
plane of the crystal) \cite{OMS}. Note that the effects described
above are caused by the carriers that are not condensed into the
CDW. Any direct effect of magnetic field on the properties of
CDWs, including the Peierls transition temperature, has never been
observed \cite{Coleman90}. In this paper, we report on the unusual
behavior of magnetoresistance of NbSe$_3$ in a magnetic field
whose orientation is perpendicular to the conducting ($b-c$)
planes.

For our study, we used high-quality NbSe$_3$ single crystals with
the ratio $R(300$K)/$R(4.2$K)$>50$. The resistance was measured by
the standard four-terminal method with a current flowing along the
chains (along the $b$ axis); the current was from 1 to 100 $\mu$A,
depending on the cross-sectional area of the sample, and, in all
of the cases, it was $2-3$ orders of magnitude smaller than the
current corresponding to the onset of the CDW slip. The magnetic
field with an induction up to 9 T was generated by a
superconducting solenoid. The measurements were performed with the
magnetic field orientation perpendicular to the ($b–c$) plane of
the crystal, while the sample could be rotated about the $c$ axis.
The temperature range of measurements was $4.2\div 60$ K.

\begin{figure}[t]
\includegraphics[width=7cm]{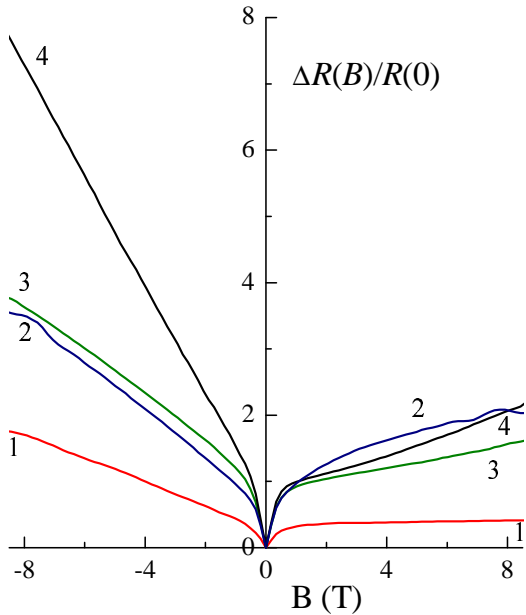}
\caption{\label{Fig1}Normalized resistance $\delta R=R(B)/R(0)-1$
versus magnetic field for four different NbSe$_3$ single crystals;
$B\parallel a^*$.}
\end{figure}

\begin{figure}[t]
\includegraphics[width=7cm]{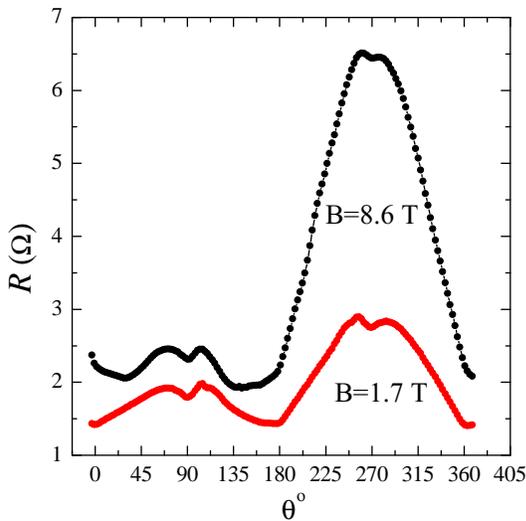}
\caption{\label{Fig2}Angular dependence of magnetoresistance for a
NbSe$_3$ single crystal (sample no. 10) rotated about the $c$ axis
at $T=4.2$ K in magnetic fields $B=8.6$ and 1.7 T (the upper and
lower curves, respectively). The angles $\theta=90^o$ and $270^o$
correspond to the magnetic field orientation parallel to the $a^*$
axis.}
\end{figure}

Figure 1 shows the normalized resistance $\delta R=R(B)/R(0)-1$
versus the magnetic field $B$ oriented along the $a^*$ axis for
four different single crystals at $T=4.2$ K. Qualitatively, the
behavior of magnetoresistance is the same for all of the samples.
In weak magnetic fields, the magnetoresistance is symmetric with
respect to the direction of magnetic field, i.e., $R(B)=R(-B)$,
and obeys the classical dependence $R\propto B^2$. In the fields
from 0.2 to 1 T, the $R(B)$ dependence changes fundamentally: from
quadratic in low magnetic fields to linear high magnetic fields.
Precisely in this field interval, starting from a certain magnetic
induction value $B_0$, the field reversal symmetry of the $\delta
R(B)$ dependences fails. The asymmetry that appears in the $\delta
R(B)$ dependences is not affected by changes in the direction and
magnitude of the transport current and is only determined by the
relative orientation of the crystal and the magnetic field. This
is illustrated by Fig. 2, which shows the angular dependence of
magnetoresistance obtained by rotating the sample about the $c$
axis for two values of magnetic field: $B=1.7$ and 8.6 T. The
presence of characteristic local maxima of magnetoresistance at
the angles $\theta=109^o$ and $\pi-109^o$ testifies to the fact
that the sample under study truly is a single crystal, because the
angle $\theta=109^o$ corresponds to the angle between the $a$ and
$c$ axes in the monoclinic crystal structure of NbSe$_3$. Note
that, to make our analysis correct, we present only the results
obtained with the samples that exhibited the aforementioned
feature, although the asymmetry under discussion was observed by
us in all other cases as well.

\begin{figure}[t]
\includegraphics[width=7cm]{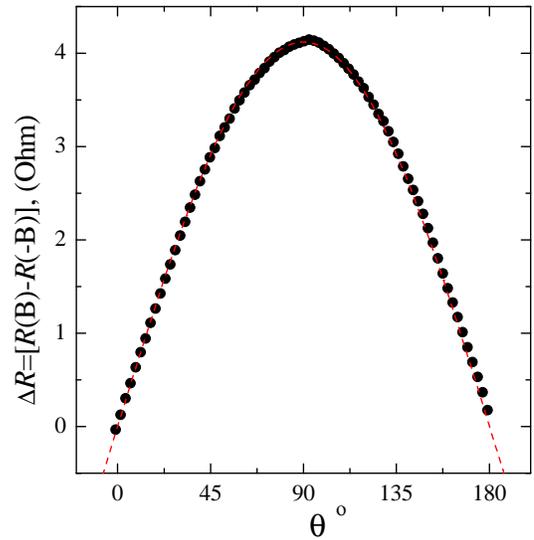}
\caption{\label{Fig3}Angular dependence of the magnetoresistance
difference $\Delta R=R(B)-R(-B)$ for a NbSe$_3$ single crystal
(sample no. 10) at $T=4.2$ K in magnetic field $B=8.6$ T. The
dashed curve represents the function $\Delta R_{max}sin\theta$.}
\end{figure}

\begin{figure}[t]
\includegraphics[width=7cm]{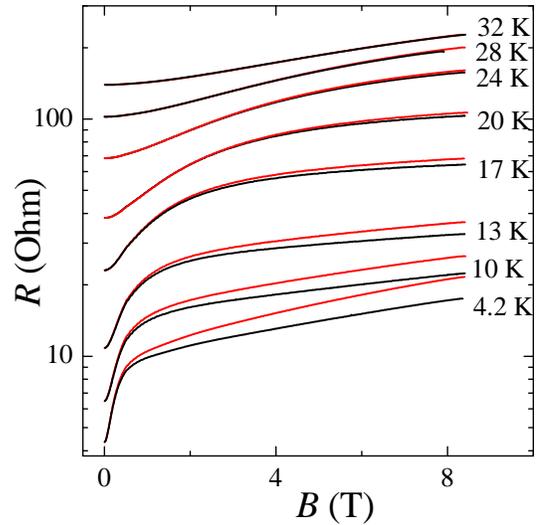}
\caption{\label{Fig4}Magnetoresistance of sample no. 13 at
different temperatures; $B\parallel a^*$.}
\end{figure}

Figure 3 displays the dependence of the difference $\Delta
R=R(B)-R(-B)$ on the angle $\theta$ at $B=8.6$ T. The experimental
dependence is adequately described by the function $\Delta
R_{max}sin\theta$ (the dashed curve in Fig. 3). This means that
only the presence of the field component parallel to the $a^*$
axis gives rise to the asymmetry of magnetoresistance. For
magnetic field orientations along the $c$ and $b$ axes, the effect
is completely absent.

The effect is also independent of history; i.e., it does not
depend on the direction of the field applied immediately after
cooling the sample to the low temperature.

Note that the presence of a similar asymmetry of magnetoresistance
in the given geometry of the experiment can be found in other
publications, for example, in Ref.\onlinecite{Monceau78}, where an
obviously asymmetric angular dependence of magnetoresistance is
presented for NbSe$_3$ in magnetic field $B=1.5$ T rotating about
the $c$ axis of the crystal. The evolution of the $R(B)$ curves
with temperature is shown in Fig. 4. At a first glance, it may
seem that, as the temperature grows, the asymmetry of
magnetoresistance decreases with the variation of the magnetic
field direction. However, one can see that the magnetic induction
$B_0$ corresponding to the appearance of asymmetry of
magnetoresistance increases with temperature. The behavior of this
parameter as a function of temperature is shown in Fig. 5. Let us
normalize the resistance by the its value $R(0)$ at $B=0$ and
normalize the magnetic induction by $B_0$. As a result, we obtain
a universal dependence shown in the inset in Fig. 5. Thus, as the
temperature increases, the effect persists, and, at high
temperatures, the asymmetry possibly arises beyond the field
interval under study. As one can see from Fig. 5, at all the
temperatures, the value of B0 falls within the region of the
qualitative change in the behavior of the $R(B)$ dependence
(deviation from quadratic dependence). At $T=4.2$ K, the value
$B_0=0.2$ T is very close to the magnetic field at which the
Landau level splitting becomes equal to temperature: $B_q=0.3$ T.
This indicates a possible quantum nature of the phenomenon under
study.

\begin{figure}[t]
\includegraphics[width=7cm]{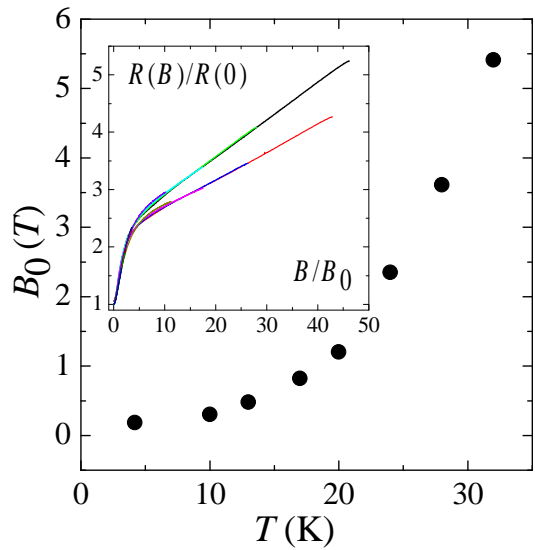}
\caption{\label{Fig5}Temperature dependence of magnetic induction
$B_0$ for the sample whose magnetoresistance is shown Fig. 4. The
inset displays the curves taken from Fig. 4 and subjected to
normalization: the magnetoresistance is normalized by $R(0)$, and
the magnetic induction, by $B_0$.}
\end{figure}

\begin{figure}[t]
\includegraphics[width=7cm]{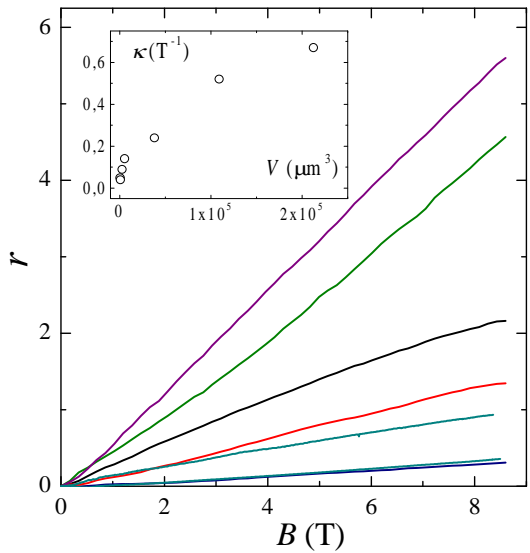}
\caption{\label{Fig6}Parameter $r=\frac{|R(+)-R(-)|}{R(0)}$ vs.
the magnetic field induction for different NbSe$_3$ samples. The
inset shows the dependence of the slope $\kappa$ of the linear
portions of the $r(B)$ dependence in high magnetic fields on the
crystal volume $V$ enclosed between the potential contacts for the
same samples.}
\end{figure}

Let us introduce a parameter to characterize the quantitative
variation of the effect. For this purpose, we use the magnetic
field dependence of the parameter:

\begin{equation}\label{E1}
      r(B)=\frac{\mid R(+B)-R(-B)\mid}{R(B=0)}
\end{equation}

As one can see from Fig. 6, which shows the behavior of this
parameter for several samples, the function $r(B)$ is linear to a
good accuracy in high magnetic fields. Hence, as a quantitative
measure of the asymmetry under observation, it is reasonable to
choose the slope, $\kappa$, of this linear dependence. We revealed
no correlation of the asymmetry with the thickness or width of the
crystals under investigation. However, from the inset in Fig. 6,
one can see that the parameter $\kappa$ monotonically increases
with increasing crystal volume enclosed between the potential
contacts, which testifies to the bulk nature of the effect.

An adequate explanation of the phenomenon described in this paper
is yet to be found. Formally, the behavior of magnetoresistance
observed in our NbSe3 samples means violation of the time reversal
invariance, which is impossible. In the quantum limit, such an
effect could be expected in the case of a spatially inhomogeneous
distribution of magnetic field formed in the sample in the
presence of local magnetic moments that may be caused by, e.g.,
magnetic impurities. However, according to the data of magnetic
susceptibility measurements \cite{Kulick79}, such impurities are
absent in NbSe$_3$. We measured the magnetic properties of
NbSe$_3$ with a high-sensitivity SQUID magnetometer in the
temperature range $4.2\div 300$ K. The data of this experiment
will be published in a separate paper. Here, we only note that
these measurements also revealed no traces of magnetic impurities
in the NbSe$_3$ single crystals.

Another possible origin of a spatially inhomogeneous distribution of
magnetic field may be the formation of toroidal magnetic moments
${\bf T}({\bf r})$ in NbSe$_3$ crystals \cite{Artamonov84},
\cite{Dubovik90}. For the case under consideration, it is important
that ${\bf T}({\bf r})$ is a polar vector, which changes sign under
time reversal. The presence of toroidal moments is allowed for 31
magnetic symmetry classes \cite{Ginzburg84}. However, NbSe$_3$ does
not belong to these kinds of magnets.

Possibly, a certain role is played by the fact that the system is in
the state with a CDW. If we consider CDW as the result of the
singlet pairing of electrons and holes, the CDW should possess no
magnetic properties. However, near the inhomogeneities of the CDW,
charge and spin density oscillations may arise, which may give rise
to local magnetic moments \cite{Slon98}. To determine the physical
mechanism of the phenomenon observed in our experiments, further
experimental and theoretical studies are necessary.

We are grateful to S.A. Brazovskiœ and V.F. Gantmakher for useful
discussions and to A.V. Kuznetsov for magnetic susceptibility
measurements. This work was supported by the Russian Foundation for
Basic Research (project nos. 05-02-17578 and 03-02-22001 CNRS) and
the INTAS (grant no. 05-7972).

\end{document}